\documentclass[preprint,12pt,sort&compress]{elsarticle}
\usepackage{amsmath,amssymb}
\usepackage{graphicx}
\usepackage{epstopdf}
\usepackage{color}
\usepackage{siunitx}

\usepackage{upgreek}

\journal{arXiv}

\begin{document}
\begin{frontmatter}

\title{Mechanical waves in myelinated axon wall}
\author{Kert Tamm, Tanel Peets, J\"uri Engelbrecht}
\address{Laboratory of Solid Mechanics, Department of Cybernetics, School of Science,\\ Tallinn University of Technology, Akadeemia tee 21, Tallinn 12618, Estonia, \\E-mails: kert@ioc.ee, tanelp@ioc.ee, je@ioc.ee}

\begin{abstract}
The propagation of an action potential is accompanied by mechanical and thermal effects. Several mathematical models explain the deformation of the unmyelinated axon wall. In this paper, the deformation of the myelinated axon wall is studied. The mathematical model is inspired by the mechanics of microstructured materials. The model involves the improved Heimburg-Jackson equation together with another equation of wave motion that describes the process in the myelin sheath. The dispersion analysis of such a model explains the behaviour of group and phase velocities. In addition, it is shown how dissipative effects may influence the process. Numerical calculations demonstrate the changes in velocities and wave profiles in the myelinated axon wall. 
\end{abstract}

\end{frontmatter}
\section{Introduction}

The propagation of signals in nerves has been studied for a long time \cite{Bishop1956,Nelson2004}. However, there are still many open questions. In the present paper, we aim to focus on one of these questions -- deformation of myelinated axon wall -- by using the mathematical modelling \cite{Raamat2021,Coveney2005,PhilArtikkel}. 

\begin{figure}[ht]
\centering
\includegraphics[width=0.66\textwidth]{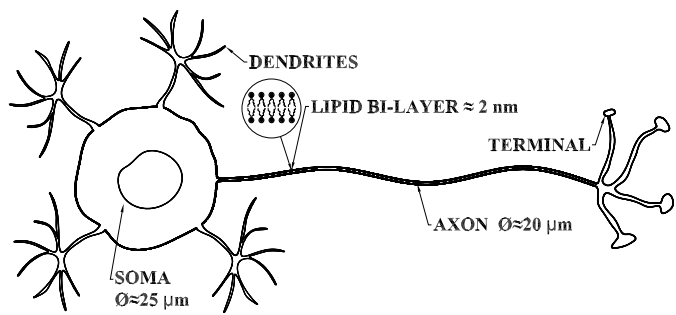}
\caption{A simplified scheme of the nerve cell.}
\label{fig0}
\end{figure}

A nerve cell (neuron) is a smallest functional unit of a nervous system. A simplified sketch of a neuron is shown in Fig.~\ref{fig0} and it consists of a cell body (also known as soma), dendrites and an axon. A nerve pulse is formed in the axon hillock (also known as the axon initial segment) \cite{Huang2018,Suminaite2019}, then it propagates along the axon to the nerve terminal (synapses) where the signal is transmitted to the next neuron. 

Axons can be either myelinated (like most mammal axons)  or unmyelinated (giant squid axon, c-fibres, certain axons in mammal brain stem and spinal cord). Unmyelinated axon walls are composed only of a single lipid bilayer together with various proteins (for example, related to ion channels). In myelinated axon walls, there are additional layers of biomembrane on the axon that are interrupted by the nodes of Ranvier. Traditionally, the myelination is viewed as a periodic structure serving as insulation.  In recent years it has been observed that the role of the myelin sheath is more than just insulation \cite{Fields2014a} and that the myeline sheath distribution along the axon is not always periodic -- the length of myeline sheaths can vary and there can be relatively long unmyelinated segments along an axon \cite{Tomassy2014}.  It is now understood that myelin changes the generation and transmission of nerve impulses and it plays an important role in neuroplasticity, learning and intelligence \cite{Hill2019}. 

It has been established by experiments that in addition to the electrical action potential (AP) a nerve signal is accompanied by mechanical and thermal components. These include the longitudinal wave in the biomembrane, the pressure wave in axoplasm and temperature changes. A mathematical model for describing the longitudinal wave in a biomembrane was proposed by Heimburg and Jackson \cite{Heimburg2005} and later improved by Engelbrecht et al. \cite{Engelbrecht2015} from the viewpoint of continuum mechanics. Here a model for longitudinal wave propagation in a myelinated axon is presented.

In Section~2, a brief overview of classical experiments is given, in Section~3, a physical description of the myelin sheath is outlined. In Section~4, a mathematical model for the mechanical wave in the axon wall is presented followed by a brief analysis of the model in Sections~5 and 6 and some numerical examples in Section~7. Section~8 contains conclusions and discussion.

\section{Classical experiments and modelling}
\subsection{Experimental observations}
The contemporary understanding of nerve signalling is based on experimental studies in unmyelinated axons. In simplified terms, axons are cylindrical tubes surrounded by extracellular fluid. The wall of the tube has a bilayered lipid structure called biomembrane formed by phospholipids (a head group and a fatty acid tail) and various proteins that all together is about 3-4~nm thick. Inside the tube is axoplasm (intracellular fluid) with cytoskeletal elements where the action potential (AP) propagates. The squid giant axon which is the classical experimental object has a diameter up to 1~mm but the usual mammalian axons have a diameter around 20~\textmu m.
A more detailed description of axon morphology is given, for example, by Debanne et al.~\cite{Debanne2011} and of biomembranes -- by Mueller and Tyler \cite{Mueller2014}. It must be noted that the structure of biomembranes is more complicated due to the various membrane proteins and ion channels. However, for the modelling of signal propagation in axons, the simple lipid bilayer description is sufficient. 

The way to understanding the propagation of APs is paved by experiments of Hodgkin and Huxley \cite{Hodgkin1952}. They have measured the AP in the unmyelinated squid axon and explained the role of ionic (Na$^+$ and K$^+$) currents. The absolute amplitude of the AP was about 100~mV, duration (without the overshoot) about 1~ms and the velocity 18.9~m/s. The contemporary measurements have demonstrated that the velocities of electric signals in nerves vary in a large interval (from ca 1~m/s to ca 100~m/s). The measurements of the AP together with the quantitative measurements of ionic currents \cite{Hodgkin1952} form the basis for the corresponding Hodgkin-Huxley (HH) model \cite{Hodgkin1964a} which is nowadays called also the Hodgkin-Huxley paradigm.

Many experiments have shown that the propagation of an AP in a nerve fibre is accompanied by transverse displacements of the biomembrane which mean changes in the axon diameter \cite{Iwasa1980,Tasaki1988,Tasaki1989}. These local changes called also swelling are small being in the range of 1-2~nm and compared with the diameter of fibres are of several orders smaller. This is confirmed by recent experiments by Yang et al.~\cite{Yang2018}. In addition to the deformation of the biomembrane, the pressure wave in axoplasm has been measured by Terakawa \cite{Terakawa1985}. In a squid axon the amplitude of a pressure wave, measured simultaneously with the AP, is about 1 to 10~mPa. The temperature change during the passage of an AP is also measured, it is of the range about 20-30~\textmu K for garfish \cite{Tasaki1988} and much less for bullfrog \cite{Tasaki1992}. The earlier findings on mechanical and thermal effects are summarised by Watanabe \cite{Watanabe1986}, and more recently by Andersen et al.~\cite{Andersen2009}.

The transverse displacement $W$ of the cylindrical biomembrane is associated with the longitudinal displacement $U$ of the biomembrane. This effect -- $W$ is proportional to the gradient of $U$ -- is well understood in mechanics \cite{Porubov2003} for the theory of rods. It means that the bipolar $W$ measured by Tasaki \cite{Tasaki1988} corresponds to the unipolar $U$ (and vice versa). The possible deformation of a biomembrane under loading is studied by measuring the transverse displacement \cite{Gonzalez-Perez2016,Perez-Camacho2017} and interpreted then as an accompanying mechanical wave along the biomembrane \cite{Heimburg2005,Gonzalez-Perez2016}. This longitudinal wave may have a soliton-type shape \cite{Heimburg2005}, i.e., is unipolar. It must be noted that the excitable plant cells (\emph{Chara braunii}) behave similarly: the electrical signal is coupled with a mechanical effect \cite{Fillafer2018}.

\subsection{Mathematical models and modelling}
\textbf{The Hodgkin--Huxley Model.}
The celebrated Hodgkin-Huxley (HH) model is a cornerstone in the contemporary understanding of axon physiology. Proposed in the mid-20th century by A.L.~Hodgkin and A.F.~Huxley \cite{Hodgkin1964a,Hodgkin1952,Schwiening2012}, this model describes explicitly the role of ion currents in forming an asymmetric AP in an unmyelinated nerve fibre.

\textbf{The FitzHugh--Nagumo model.} 
The existence of many physical parameters in the HH model (see above) may cause problems in modelling. That is why the simplified models are sought that still could describe the main effects. For example, FitzHugh \cite{FitzHugh1961} used the ideas of Bonhoeffer \cite{Bonhoeffer1948} and van der Pol \cite{vanderpol1926} for deriving a model of an excitable-oscillatory system. FitzHugh named his model after Bonhoffer and van der Pol but starting from the paper by Nagumo et al. \cite{Nagumo1962}, the model is called after FitzHugh and Nagumo.

\textbf{The evolution equation (the Engelbrecht model).} 
From wave mechanics, it is known that the classical wave equation describes two waves -- one propagating to the right, another -- to the left. Under certain conditions, these waves can be separated and in this case, the result is an evolution equation that describes one wave -- either propagating to the right or the left. The evolution equation for the nerve pulse has been derived by Engelbrecht \cite{Engelbrecht1981}. In mathematical terms, the leading derivative is then of the first order. The details of such derivation by the reductive perturbation method are described in several monographs \cite{Engelbrecht1983,taniuti1983}.

\textbf{The coupled model.}
There are the mechanical effects as can be seen from the classical experimental works noted in the previous subsection. The propagation of the AP is accompanied also by the dynamical deformation of the biomembrane. A mathematical model for the longitudinal deformation is proposed by Heimburg and Jackson \cite{Heimburg2005} and later improved by Engelbrecht et al.~\cite{Engelbrecht2015,Engelbrecht2017,Peets2019a,Peets2019}.

A model framework is recently published by Engelbrecht et al.~in a monograph \cite{Raamat2021} giving a systematic overview of modelling the nerve pulse and accompanying effects in non-myelinated axons. It is beneficial to enlarge the ideas presented in  \cite{Raamat2021} to myelinated axons. As before, in order to construct a full model, one should understand the role of structural elements of an axon in this case. This is the model we are following in the present paper. 

\textbf{The mechanoelectrical model} An alternative model for coupling mechanical and electrical aspects of the nerve pulses has been proposed which ties the membrane surface potential changes to the changes in the membrane curvature \cite{Fillafer2018,Chen2019,Mussel2019,Jerusalem2019}. Briefly, the idea is that as the lipid bi-layer curvature changes (if there is a mechanical wave) the membrane surface potential changes because the lipid molecules are asymmetrically charged. Similarly, as the membrane surface potential changes (if there is an electrical signal) the membrane curvature should change as well. We draw some inspiration from the ``flexoelectric" effects described in the ``mechanoelectrical" model for proposing a speculative mechanism of how mechanical waves in myelin sheath could be responsible for speeding up the AP in the myelinated axons later in the paper. 

\section{Physical description}

The axon can be described as a tube filled with axoplasm and cytoskeleton with a lipid-bilayer (biomembrane) wall which is embedded in an intercellular medium (fluid). While some axons are unmyelinated, many of them have a myelin sheath that is in simplified terms composed of layers of biomembrane glued together with some proteins. The myelin sheath is interrupted by the nodes of Ranvier which play important role in the nerve pulse propagation. A node of Ranvier is typically around 1~\textmu m in length and has a high density of ion channels \cite{Lodish2004} while myelin sheath segments are typically from roughly 50~\textmu m to 300~\textmu m in length. The distribution of the myelinated parts was once believed to be uniform, but now it is understood that the length of the myelinated parts and the nodes of Ranvier vary and unusually long nodes of Ranvier (50 $\upmu$m) have been reported \cite{Tomassy2014}. These long segments could play important role in the synchronisation of nerve pulses by delaying an AP. 

The myelin sheath is a stack of specialised plasma membrane sheets produced by glial cells that wrap around the axon \cite{Lodish2004}. The myelin sheath is characterised by a high proportion of lipids (70\%–85\%) and consequently, a low proportion of proteins (15\%–30\%) \cite{Poitelon2020}, while in most biological cells this ratio is 50\% lipid/50\% protein. 
 A lipid bilayer (biomembrane) is typically from 3 to 4 nm in thickness and the main protein P$_0$ (acting as a glue between lipid bi-layers) in the myelin sheath has about 4.5 nm central part and roughly similar size `tails' embedded in adjacent lipid bi-layers \cite{Raasakka2019} that all together form a myelin sheath composed by many such layers surrounding the axon. The myelin sheath can be up to 2.5~\textmu m in thickness \cite{Sanders1948,Michailov2004,Hanig2018} and on the lowest limit, an extra layer of biomembrane surrounding the axon might be argued to be a myelin sheath. Typical axon diameter varies from about 0.5~\textmu m to 25~\textmu m \cite{Sanders1948} in mammals but can reach as high as about 1~mm in giant squids \cite{Hodgkin1949,Terakawa1985}.

\begin{figure}[ht]
\includegraphics[width=0.99\textwidth]{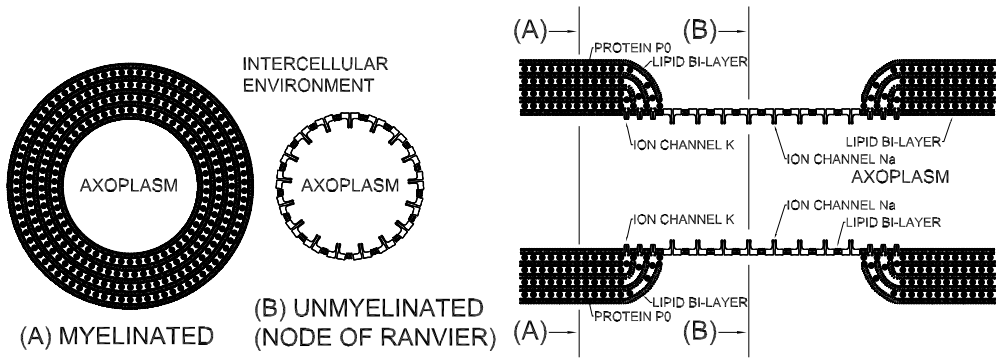}
\caption{A simplified scheme of the geometry of the myelinated axon. (A) a cross-section of the myelinated axon segment, (B) a cross-section of the unmyelinated section of the axon (node of Ranvier).}
\label{fig1}
\end{figure}

Nodes of Ranvier contain much higher densities of various ion channels than elsewhere on axons. K and Ca channels are about 10~nm in length and roughly 4~nm in diameter \cite{Doyle1998}. Na channel is about 12~nm long and about 10nm in diameter \cite{Sula2017}. The myelinated axon geometry is sketched in Fig.~\ref{fig1}.

In the peripheral nervous system (PNS), the myelination is produced by the Schwann cells and it will be completed shortly after birth (within two years after birth in humans; within four weeks after birth in rodents) \cite{Poitelon2020}. The myelination in the central nervous system (CNS) is a lifelong process and is formed by the oligodendrocytes. One oligodendrocyte myelinates several neurons (up to 30 and more), but Schwann cell only myelinates one segment for a single neuron \cite{Poitelon2020,Williamson2018,Salzer2016}.

Structurally the myelinated part of the axon is divided into the following regions: next to the node of Ranvier is a region called paranode. This is the area where the myeline attaches to the axon. Juxtaparanode is located next to the paranode and it is the area where most voltage-gated K$^+$ ion channels are located. The Na$^+$ channels are concentrated in the nodes of Ranvier.

The AP propagates down an axon
without diminution at speeds up to 1~m/s (without myelin sheath) \cite{Lodish2004}. In non-myelinated neurons, the conduction velocity of an action potential is roughly proportional to the diameter of the axon \cite{Lodish2004}. The presence of a myelin sheath around an axon increases the velocity of impulse conduction to 10-100~m/s \cite{Lodish2004}.


\section{Model formulation}

Here we present a model for an elastic longitudinal wave in myelinated axon wall by taking inspiration from the theory of microstructured solids \cite{Mindlin1964,Berezovski2010a,Berezovski2013}. 
We stress that modelling of such a complicated inhomogeneous structure as described in Section 3 is dependent on many physiological properties. However, here we aim to grasp the essential effects due to the myelin sheath in the proposed model. 
The starting point of the modelling is the improved Heimburg-Jackson (HJ) model for the longitudinal mechanical wave in the lipid bilayer (the homogeneous lipid bilayer without ion channels) \cite{HJ2007,Engelbrecht2018,Engelbrecht2020m}
\begin{equation} \label{ihjeq}
\begin{split}
U_{TT} &=  c_{0}^{2} U_{XX} + N U U_{XX} + M U^2 U_{XX} +
   N U_{X}^{2} + 2 M U U_{X}^{2}-\\
   & H_1 U_{XXXX} + H_2 U_{XXTT} - \mu U_T + F(Z,J,P),
\end{split}
\end{equation}
where $U = \Delta \rho$ is the longitudinal density change, $c_{0}$ is the sound velocity in the unperturbed state, $N$,$M$ are nonlinear coefficients, $H_i$ ($i=1,2$) are dispersion coefficients and $\mu$ is the dissipation coefficient. Note that the dispersive coefficients $H_i$ model the influence of the microstructure of the lipid bilayer --  $H_1$ accounts for the elastic properties and $H_2$ -- for the inertial properties. Here $F$ is the coupling force accounting for the possible influence from the AP and PW. In $F(Z,J,P)$ the $Z$ represents action potential, $J$ the ion current(s) and $P$ the pressure change in axoplasm. The experimentally measured transverse displacement (W) of the biomembrane (TW) is taken as in the theory of rods  $W \propto U_X$ \cite{EngelbrechtTammPeets2014}. 

Originally, the Heimburg-Jackson model is formulated based on experimental considerations \cite{Heimburg2005,HJ2007} and is a Boussinesq-type equation \cite{Christov2007}. However, one option of reaching that type of equations of motion mathematically is through the Lagrange formalism, which, very briefly, means constructing a Lagrangian $L=K-W$, where $K$ is kinetic energy and $W$ is free (potential) energy \cite{Arkadi2006,Peets2015}. It should be noted that using Lagrange formalism means assuming that we have a conservative system. Then, equations of motion are derived from
\begin{equation} \label{lagrange}
\left(\frac{\partial L}{\partial u_t}\right)_{t} + \left(\frac{\partial L}{\partial u_x}\right)_x - \frac{\partial L}{\partial u} = 0,
\end{equation} 
where $u$ is displacement, $\partial / \partial \ldots$ denotes partial derivative and subscripts $x$, $t$ represent a partial derivative with respect to space and time, respectively. This remark on Lagrangian formalism is interesting because of somewhat uncommon nonlinear terms in the HJ model. Normally in continuum mechanics, the nonlinear terms in the Boussinesq-type equations are formulated by gradients. To get the nonlinear terms similar to the improved HJ model \eqref{ihjeq} through the Lagrangian formalism, the free energy would need to include the terms $N  u u_{x}^{2}$ and $M  u^2 u_{x}^{2}$ whose physical interpretation is an open question \cite{Peets2015}. In reality, the mechanical component of the nerve pulse is not propagating in a conservative system, as is evident from the dissipative and coupling force terms in Eq.~\eqref{ihjeq}. 

The improved HJ model~\eqref{ihjeq} is derived for an unmyelinated axon and it involves only one microstructure -- lipid structure which is modelled by the dispersive terms $U_{XXXX}$ and $U_{XXTT}$. Propagation of the AP in the myelinated axon is influenced by the thickness of the myelin structure which is characterised by the g-ratio (the ratio between the axon radius to radius with myelination) and by the nodal and internodal lengths \cite{Mohammadi2015,Schmidt2019}. For modelling the mechanical wave propagation in myelinated axons, these effects and possible inhomogeneities  (different ion channel densities, for example, \cite{Suminaite2019,Arancibia-Carcamo2014}) could be relevant. 

\begin{figure}[!h!t]
\includegraphics[width=0.467\textwidth]{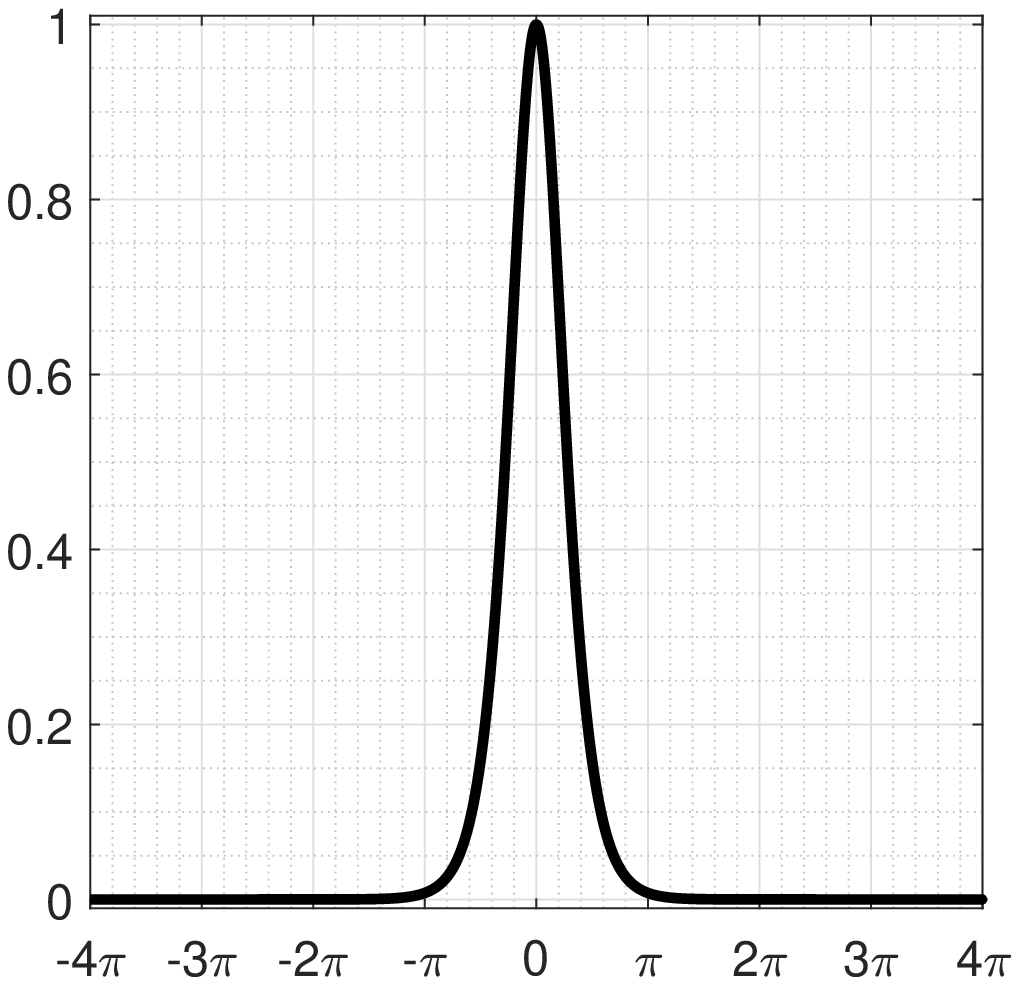}
\includegraphics[width=0.49\textwidth]{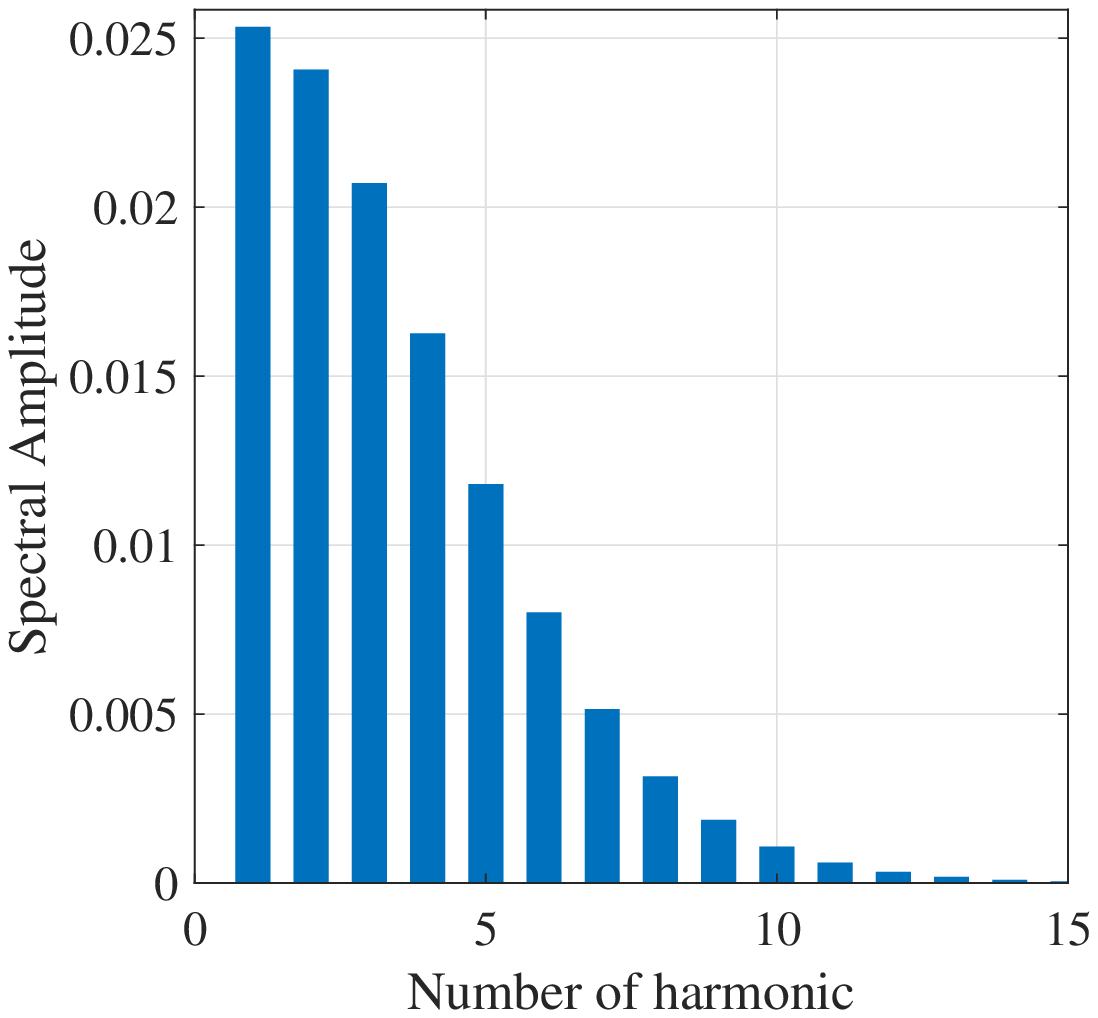}
\caption{The left panel -- $\mathrm{sech}^2(x)$ profile, the right panel -- corresponding spectral amplitudes \cite{salup2009} for such a $\mathrm{sech}^2$ profile. }
\label{fig2}
\end{figure}

The important question is the scale of the problem. Typical nerve pulse wavelengths range from about 20~cm (fast, long-lasting pulse)  to roughly 4~mm (slow, short-duration pulse). Here, the wavelength of a signal is taken as a distance from the front of the pulse until the pulse amplitude returns close to the equilibrium value at a fixed time. This is several orders of magnitude longer than the structure of the elements of the myelinated axon (myelin sheath is from 50 to 300~\textmu m and gaps between these roughly 1~\textmu m). In Fig.~\ref{fig2}  the spectral composition \cite{salup2009} of a typical bell-shaped pulse is shown where it can be seen that while the bulk of the signal is contained in lower harmonics, there is a sufficient number of higher harmonics whose wavelengths can be of the same order of magnitude as nodes of Ranvier as the wavelength of harmonics decreases rapidly for the higher harmonics.

Since the myelinated axon can be viewed as a microstructured material then  inspiration can be drawn from the Mindlin-type model for materials with double microstructure  \cite{Arkadi2006,Peets2013,Tamm2016}:
\begin{equation}\label{ConcDFSE}
\begin{split}
& \rho U_{TT} = \alpha U_{XX} + A_1 (\Phi_1)_X + A_2 (\Phi_2)_X + N  U_{X} U_{XX},
\\
& I_1  (\Phi_1)_{TT} = C_1 (\Phi_1)_{XX}  - B_1  \Phi_1 - A_1  U_X  + M_1  (\Phi_1)_{	X} (\Phi_1)_{XX},
\\
& I_2  (\Phi_2)_{TT} = C_2 (\Phi_2)_{XX} - B_2  \Phi_2 - A_2  U_X  + M_2  (\Phi_2)_{X} (\Phi_2)_{XX}  .
\end{split}
\end{equation}
Here $U$ is displacement, $\Phi_{i}$ are microdeformations, $\alpha, A_i,B_i,C_i,N_i,M_i$ are parameters, $\rho$ is density and $I_i$ are microinertias (here $i=1,2$). The indices $X,T$ denote differentiation with respect to space and time, respectively. A very brief explanation of Eqs.~\eqref{ConcDFSE} is that these model a mechanical wave $U$ in 1D setting which is affected by several microstructures $\Phi_1, \Phi_2$ whose mathematical description is, in essence, a ``field" which interacts with the macro-scale mechanical wave through coupling $A_1,A_2$. 

System~\eqref{ConcDFSE} models a concurrent model where two microstructures interact with a macrostructure. It is also possible to consider a hierarchical model where microstructure $\Phi_2$ interacts with microstructure $\Phi_1$ which in turn interacts with the macrostructure (a scale within a scale). For a mechanical wave in myelinated axon wall, the concurrent model is preferred as the structure of a biomembrane and myeline sheath spatial distribution along the axon can be treated as two separate microstructures. 

\begin{figure}[ht]
\includegraphics[width=0.99\textwidth]{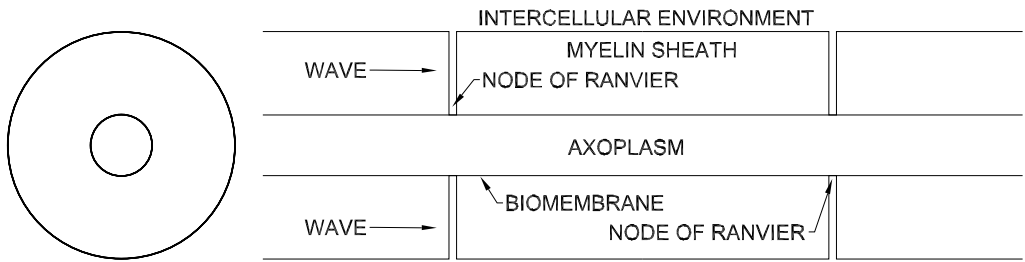}
\caption{A sketch of the model environment for mechanical waves in the myelinated axon. The axon is taken as a tube with the variable wall thickness filled with axoplasm and surrounded by an intercellular environment. The myelin and biomembrane are taken with identical mechanical properties. The tube sections with thicker walls represent myelinated sections and the gaps between these represent nodes of Ranvier. In such a tube a longitudinal mechanical wave propagates with a higher wavelength than the gap between the myelinated sections. The density of axoplasm and the intercellular medium is roughly the same as the density of lipids meaning that the mechanical wave can propagate through the gaps without major disturbances. Some energy of the mechanical wave could be transferred to the axoplasm and into the intercellular medium.}
\label{fig3}
\end{figure}

Considering the scales of the problem it can be argued that we can omit the second (smaller) microstructure layer, as too small in scale to significantly affect the mechanical wave in the axon wall. The first harmonics of the mechanical signal are from about 20 cm to 4 mm, somewhere around the 12th harmonic we get into the $\upmu$m  range which we consider as the scale of our first microstructure layer (nodes of Ranvier). The next smaller scale (which we omit here) would be the influence of the ion channels embedded within the nodes of Ranvier on the mechanical longitudinal density wave which is a few orders of magnitude smaller at roughly 10 nm in size.
The geometry of the model problem is depicted in Fig.~\ref{fig3}. 

System~\eqref{ConcDFSE} includes three coupled equations, but it has been shown \cite{Arkadi2006} that the coupled system can be expressed as one Boussinesq-type equation by making use of the slaving principle. Taking inspiration from Mindlin-type \eqref{ConcDFSE} and improved HJ models, the following system is proposed for the mechanical wave propagation in the myelinated axon wall:
\begin{equation} \label{iHJ2}
\begin{split}
U_{TT} &=  \gamma_{0}^{2} U_{XX} + N U U_{XX} + M U^2 U_{XX} +
   N U_{X}^{2} + 2 M U U_{X}^{2}-\\
   & H_1 U_{XXXX} + H_2 U_{XXTT} - \mu U_T + F(Z,J,P) + A_1 \Phi_{X},\\
\Phi_{TT} &= \gamma_2^2 \Phi_{XX} - \eta_0^2 \Phi - A_2 U_X,
\end{split}
\end{equation}
where  $\Phi$ models the influence of the myelination on the macroscopic longitudinal wave propagation; $\gamma_0$, $\gamma_2$ and $\eta_0$ are dimensionless characteristic velocities and frequency, $A_i$ are coupling constants. 
A possible need to model the effect of additional microstructures or inclusion of higher frequency components can be satisfied by adding an additional microstructure, i.e., by taking inspiration from system~\eqref{ConcDFSE}. It should be noted that it is possible to write system \eqref{iHJ2} as a single equation by making use of the slaving principle \cite{ebpb2005}, which enrols the influence of the microstructure into the macro-scale description of the wave process. In other terms, it means the appearance of the 6th order dispersion terms ($U_{XXXXXX}$ and $U_{TTXXXX}$) in the governing equation. Here, the preference is to keep $\Phi$ separate to emphasise that $\Phi$ is an internal variable \cite{Van2008,Berezovski2011a}. Mechanically $\Phi$ introduces an additional degree of freedom into the system and allows a second (higher-order, optical) dispersion branch. Naming the higher-order dispersion branches `optical' and the lowest branch `acoustical' is the historical nomenclature.
\begin{figure}[!h]
\centering
\includegraphics[width=0.75\textwidth]{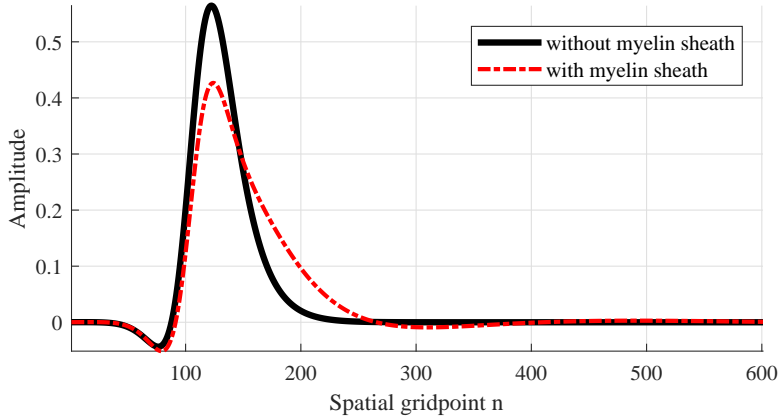}
\caption{An example solution of a conservative iHJ equation (dissipation and coupling force omitted) compared to the iHJ equation where myelin sheatch has been included (Eqs.~\eqref{iHJ2}) as a microstructure. In Eqs.~\eqref{iHJ2} the parameters are the following: $\gamma_{0}^{2}=1$, $N=0.01$, $M=0.0005$, $H_1=0.205$, $H_2=0.2$, $\mu=0$ (no dissipation), $F(Z,J,P)=0$ (no coupling force), $\gamma_{2}^{2}=\gamma_{0}^{2}$, $\eta_{0}^{2}=0.05$, $A_2=0.05$ and $A_1=0$ (without myelin case) or $A_1=0.01$ (with myelin case). The initial condition is a bell shaped profile in the middle of spatial period splitting into two counter-propagating bell shaped pulses. The pulses propagating to the left are shown at $T=100$.}
\label{fig4}
\end{figure}
An example solution of system~\eqref{iHJ2} is demonstrated in Fig.~\ref{fig4}. 

It should be stressed that the Mindlin-type microstructured materials \cite{Mindlin1964,Berezovski2013} are described within the framework of continuum mechanics and the whole process is formulated from the viewpoint of the macroscale. It means that the additional equations included for the microstructures describe the influence of the microstructure on the wave processes on the macroscale. In other words, the microstructure is represented as a field that modifies wave behaviour at a macroscopic (measurable/observable) scale. In practical terms, each microstructure scale adds another degree of freedom into the wave process, which, in turn, means additional modes of wave propagation \cite{Peets2016MatCompSim}.

\section{Dispersion analysis}

For dispersion analysis the system~\eqref{iHJ2} with nonlinear, dissipative and forcing terms omitted will be used:
\begin{equation} \label{iHJ2-linear}
\begin{split}
U_{TT} &=  \gamma_{0}^{2} U_{XX} - H_1 U_{XXXX} + H_2 U_{XXTT} + A_1 \Phi_{X},\\
\Phi_{TT} &= \gamma_2^2 \Phi_{XX} - \eta_0^2 \Phi - A_2 U_X.
\end{split}
\end{equation}
It is possible to rewrite the system~\eqref{iHJ2-linear} as one PDE. To that end the second equation in system~\eqref{iHJ2-linear} is differentiated once against $X$; the second equation is solved for $\Phi_X$ and is substituted into the second equation. The following 6th order PDE is obtained:
\begin{equation}
	\label{onePDE}
	\begin{split}
	U_{TT}=&(\gamma_0^2-\frac{A_1A_2}{\eta_0^2})U_{XX}+H_2(U_{TT}-\frac{H_1}{H_2}U_{XX})_{XX}\\
	&-\frac{1}{\eta_0^2}\left[(U_{TT}-\gamma_0^2U_{XX})_{TT}-\gamma_2^2(U_{TT}-\gamma_0^2U_{XX})_{XX}\right]\\
	&+\frac{H_2}{\eta_0^2}\left[\left(U_{TT}-\frac{H_1}{H_2}U_{XX}\right)_{TT}-\gamma_2^2\left(U_{TT}-\frac{H_1}{H_2}U_{XX}\right)_{XX}\right]_{XX}
	\end{split}
\end{equation}
Equation~\eqref{onePDE} demonstrates the hierarchical nature of wave propagation in coupled systems. The dimensionless parameter $\eta_0$ can be viewed as a characteristic frequency \cite{Berezovski2010a} that controls the behaviour of shorter wavelength (high frequency) harmonics. For long wavelengths, the wave propagation is dominated by the first two terms on the rhs of Eq.~\eqref{onePDE}.
The velocity of low-frequency harmonics is reduced due to the coupling with the field $\Phi$.
As the wavelengths get shorter, the wave propagation becomes first dominated by the operator  $U_{TT}-\gamma_0^2U_{XX}$ and then in the short wave limit by the operator $U_{TT}-\gamma_1^2U_{XX}$, where $\gamma_1^2=H_1/H_2$.

To study the dispersion of wave propagation governed by the system~\eqref{iHJ2-linear} we assume the solutions in the form of harmonic waves:
\begin{equation}
\label{DA1}
U(X,T)=\hat{U}e^{i(kX-\omega T)}, \quad \Phi(X,T)=\hat{\Phi}e^{i(kX-\omega T)}, 
\end{equation}
where $k$ and $\omega$ are the wave number and frequency respectively.  Plugging these into system~\eqref{iHJ2-linear} we get
\begin{equation}
\label{matrixDISP}
\left|\begin{array}{cc}\omega^2-k^2\gamma_0^2-H_1k^4+H_2k^2\omega^2 & iA_1k \\-iA_2k & \omega^2-k^2\gamma_2^2-\omega_0^2 \end{array}\right| \left|\begin{array}{c}\hat{U} \\\hat{\Phi} 
\end{array}\right|=0.
\end{equation}
In order to get nontrivial solutions, the determinant of this system must vanish. This leads to the dispersion relation
\begin{equation}
	(\omega^2-k^2\gamma_0^2-H_1k^4+H_2k^2\omega^2)(\omega^2-k^2\gamma_2^2-\eta_0^2)-A_1A_2k^2=0,
\end{equation}
which can be rewritten as 
\begin{equation}
	\label{disprel}
	\begin{split}
	\omega^2=&\left(\gamma_0^2-\frac{A_1A_2}{\eta_0^2}\right)k^2-H_2(\omega^2-\gamma_1^2k^2)k^2\\
	&+\frac{1}{\eta_0^2}(\omega^2-\gamma_0^2k^2)(\omega^2-\gamma_2^2k^2)
	-\frac{H_2}{\eta_0^2}(\omega^2-\gamma_1^2k^2)(\omega^2-\gamma_2^2k^2)k^2.
	\end{split}
\end{equation}
In Eq.~\eqref{disprel}, the asymptotic velocities can be clearly seen -- in the long-wave limit these velocities are $(\gamma_0^2-A_1A_2/\eta_0^2)^{1/2}$ and $\gamma_1=(H_1/H_2)^{1/2}$. Note that due to the coupling in system~\eqref{iHJ2-linear}, the velocity $\gamma_0^2$ is reduced by a factor of $A_1A_2/\eta_0^2$.   In the medium wave limit, three characteristic velocities exist: $\gamma_1=(H_1/H_2)^{1/2}$, $\gamma_0$ and $\gamma_2$ and in the short wave limit, the asymptotic velocities are $\gamma_0$ and $\gamma_2$. The influence of the field $\Phi$ is controlled by the parameter $\eta_0$ -- when $\eta_0$ becomes large then the correction to velocity $\gamma_0$ and the higher-order terms vanish.  

\begin{figure}
	\centering
	\includegraphics[width=.49\textwidth]{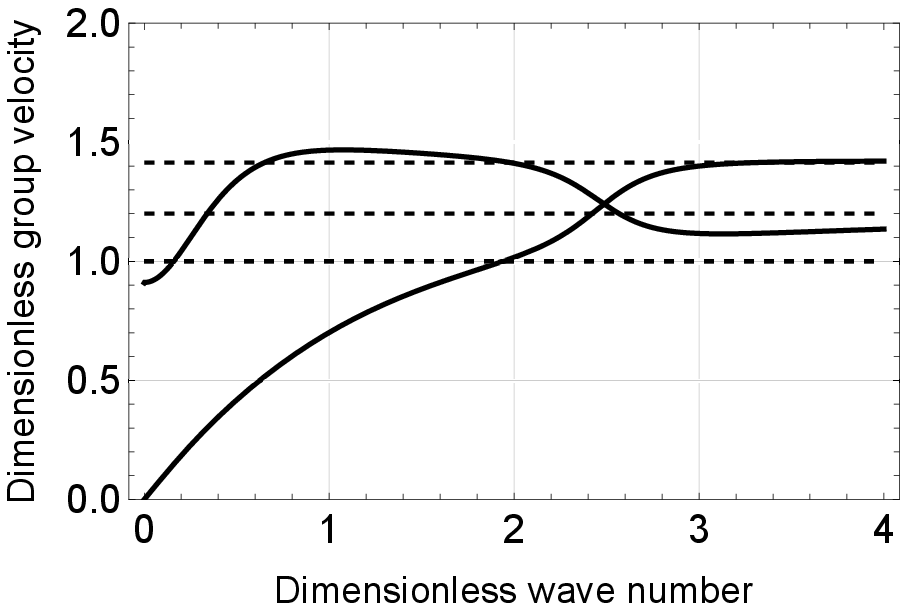}
	\includegraphics[width=.49\textwidth]{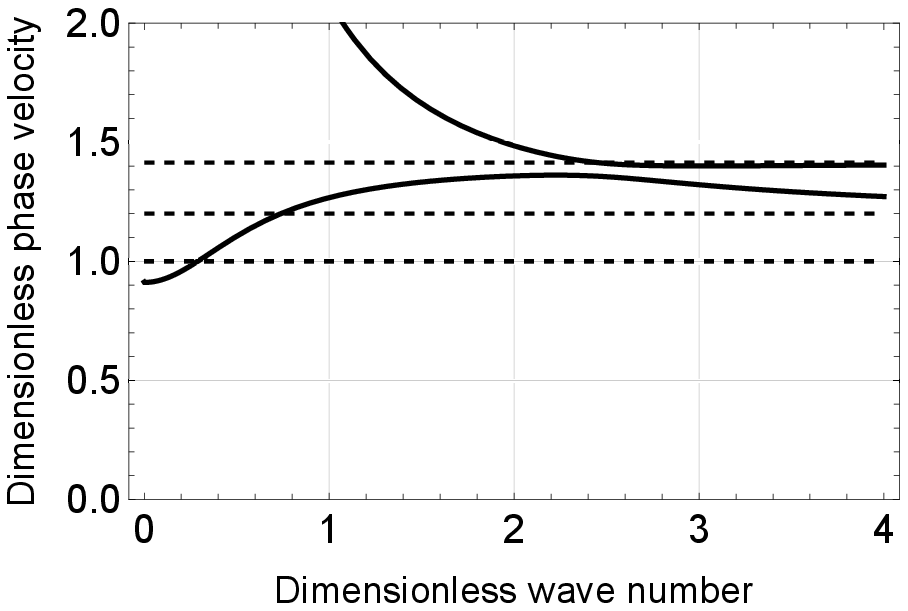}
	\caption{Group (the left panel) and phase speed (the right panel) curves for  the case of $\gamma_0=1$, $H_1=4$, $H_2=2$ ($\gamma_1\approx 1.4$), $\gamma_2=1.2$, $\eta_0=1.7$ and $A_1=A_2=0.7$.  Dashed horizontal lines represent characteristic velocities -- $\gamma_0$ (bottom), $\gamma_2$ (middle) and $\gamma_1=(H_1/H_2)^{1/2}$ (upper). }
	\label{GroupSpeed}
\end{figure}

The asymptotic behaviour of the dispersion relation is visible in Fig.~\ref{GroupSpeed} where the group ($c_g=\partial \omega/\partial k$) and phase ($\omega /k$)  velocity  curves are plotted. Due to the coupling in Eq.~\eqref{iHJ2-linear} two distinct branches exist. The acoustic branch starts with velocity $(\gamma_0^2-A_1A_2/\eta_0)^{1/2}$ and then approaches the velocity $\gamma_1$ and then $\gamma_2$. The optical curve starts with zero group velocity (or infinite phase velocity) and approaches the velocity $\gamma_2$ and then to velocity $\gamma  _1$ in the short wave limit. It must be stressed that the parameters in Fig.~\ref{GroupSpeed} have been chosen in such a way that the behaviour of group and phase speed curves would be clearly visible. The physiologically viable parameters should be determined from experiments. 

\section{Dissipation}

The improved HJ equation with dissipation (first equation in system~\eqref{iHJ2-linear}) includes first-order dissipative term $-\mu U_T$. Due to the coupling in system~\eqref{iHJ2-linear} higher order dissipative terms arise. This effect is similar to dispersion -- the improved HJ equation only includes 4th order dispersive terms, but due to coupling 6th order terms arise. To demonstrate this,  system~\eqref{iHJ2} without nonlinear terms is expressed as one PDE using the technique described in the previous subsection:
\begin{equation}
	\label{onePDEWithDissip}
	\begin{split}
	&U_{TT}=(\gamma_0^2-\frac{A_1A_2}{\eta_0^2})U_{XX}+H_2(U_{TT}-\frac{H_1}{H_2}U_{XX})_{XX}-\mu U_T \\
	&-\frac{1}{\eta_0^2}\left[(U_{TT}-\gamma_0^2U_{XX}+\mu U_T)_{TT}-\gamma_2^2(U_{TT}-\gamma_0^2U_{XX}+\mu U_T)_{XX}\right]\\
	&+\frac{H_2}{\eta_0^2}\left[\left(U_{TT}-\frac{H_1}{H_2}U_{XX}\right)_{TT}-\gamma_2^2\left(U_{TT}-\frac{H_1}{H_2}U_{XX}\right)_{XX}\right]_{XX}.
	\end{split}
\end{equation}
The validity of the derived equation can be confirmed with dispersion analysis by finding the dispersion relation from linearised system~\eqref{iHJ2} and comparing it to the dispersion relation obtained from Eq.~\eqref{onePDEWithDissip}. In both cases three additional terms arise -- $-i\mu\omega$, $-i\mu (\gamma_2/\eta_0)^2\omega k^2$ and $i/\eta_0^2\omega^3$.

An important question is the existence of dissipative members in Eqs.~\eqref{iHJ2} as the model geometry is altered compared to the situation for which Eq.~\eqref{ihjeq} was intended to describe. In Fig.~\ref{fig3} we have, in essence, an 1D problem for a longitudinal mechanical wave in the model tube. However, as in rods or tubes, the longitudinal wave is coupled to the transverse displacement of the boundary layers (see, for example, \cite{Porubov2003}) some of the mechanical energy can be transferred into the axoplasm and intercellular medium. For the mechanical longitudinal wave in the axon wall, any such energy is, essentially, lost, meaning it could be modelled as some kind of additional dissipative element in Eqs.~\eqref{iHJ2} (i.e., in the simplest case just splitting the dissipation coefficient $\mu=\mu_1+\mu_2$). It would be prudent to differentiate between the `classical' dissipative processes which convert some mechanical energy into heat through friction/viscosity type processes (parameter $\mu_1$) and a process that just represents a loss of energy into the surrounding medium (parameter $\mu_2$, which is taken as zero in the present paper). In this sense, the model could be called 1.5D where the mathematical description in 1D but some interaction with the surrounding medium is taken into account indirectly without explicitly modelling the processes perpendicular to the axis of the axon.

\section{Numerical examples}
The pseudospectral method (PSM) (see \cite{Fornberg1998,salup2009,Raamat2021}) is used to solve the system of dimensionless model equations~\eqref{iHJ2}. A $\mathrm{sech}^{2}$-type localised initial condition with initial amplitudes $A_o$ and width $B_o$ is used in the middle of the spatial period and the periodic boundary conditions for all model equations are taken into account. The idea of the PSM is to approximate space derivatives by making use of the properties of the Fourier transform reducing, therefore, the partial differential equation (PDE) to an ordinary differential equation (ODE) and then to use standard ODE solvers for integration in time. Normally the PSM algorithm is intended for $ u_t = \Phi(u,u_x, u_{2x},\ldots,u_{mx})$ type equations. However, we have a mixed partial derivative term $H_2 U_{XXTT}$ and as a result some modifications are needed (see \cite{lauriandrus2009,lauriandruspearu2007,salup2009}). The detailed description of the numerical scheme used can be found in \cite{Raamat2021} Appendix A and an example code (written in Python) demonstrating the application of the used numerical scheme for a system including the improved Heimburg-Jackson equation can be found in Appendix B in \cite{Raamat2021}.

For the initial conditions, we use a $\mathrm{sech}^{2}$-shaped pulse in the middle of the spatial period with initial amplitude of $A_o = 1$ and width parameter $B_o = 0.05$ with zero initial velocity (i.e., $U(X,0)_T = 0$) which means the initial pulse will separate into two counter-propagating localised pulses with amplitude of 0.5. For $\Phi$ we set initially $\Phi(X,0) = \Phi(X,0)_T = 0$, i.e., the system is at rest at $T=0$. The length of the spatial period is taken $128\pi$ and the number of spatial grid points is set at $n=8192$. In the following figures, the profiles propagating to the left are shown. 

\begin{figure}[h]
\includegraphics[width=0.99\textwidth]{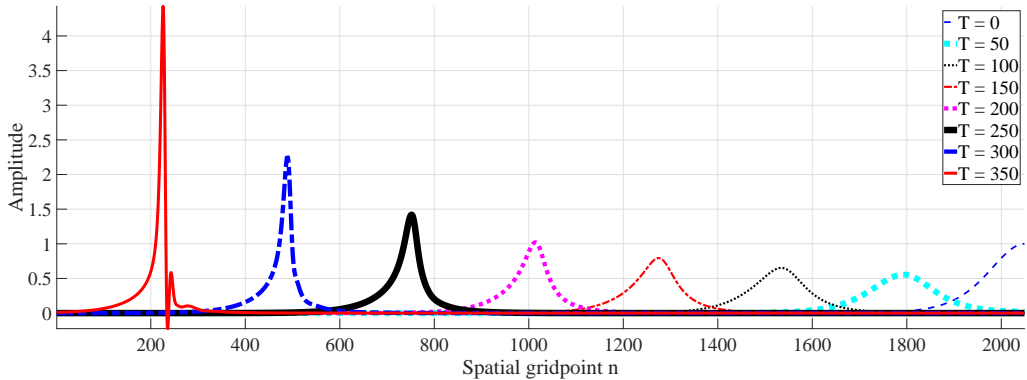}
\caption{The time evolution of the left propagating pulse in the half-space. The initial pulse with an amplitude of 1 splits into two counter-propagating pulses with an amplitude of 0.5 (only the left half-space is plotted). During the propagation, the pulse shape changes into a more narrow one and the amplitude increases until eventually, the localised pulse breaks apart into an oscillatory package (T=350 and beyond).}
\label{fig5}
\end{figure}

The parameters for system~\eqref{iHJ2} are taken as $\gamma_0^{2}=1$, $N=0.001$, $M=0.0005$, $H_1=0.2$, $H_2=0.20001$, $\mu = 0$ (no dissipation), $F(Z,J,P)=0$ (only the mechanical wave without the driving force) and the influence of the parameters $\gamma_1^2$, $\eta_0^2$, $A_i$ is investigated. A `reference' case against which the influence of the parameters $\gamma_1^2$, $\eta_0^2$, $A_i$ is compared is $\gamma_2^2 = \gamma_1^2 = 1$, $\eta_0^2 = 0.05$, $A_1 = 0.005$, $A_2 = 0.1$. We compare the numerical solutions at $T=250$. The evolution of the left-propagating pulse is shown in Fig.~\ref{fig5}. With these parameter values, the acoustic dispersion branch is almost straight line which means a balanced dispersion case (i.e., all frequencies propagate with roughly equal velocity). Due to the existence of the optical dispersion branch, which reflects the effect of an additional field $\Phi$,  the wave propagation is dispersive. 
At the same time, we have nonlinearity in the system which can redistribute energy between the harmonics of the signal, shifting some energy from the lower frequencies towards the higher frequencies. For these reasons the wave profile propagating to the left is deformed. 

\begin{figure}[h]
\includegraphics[width=0.49\textwidth]{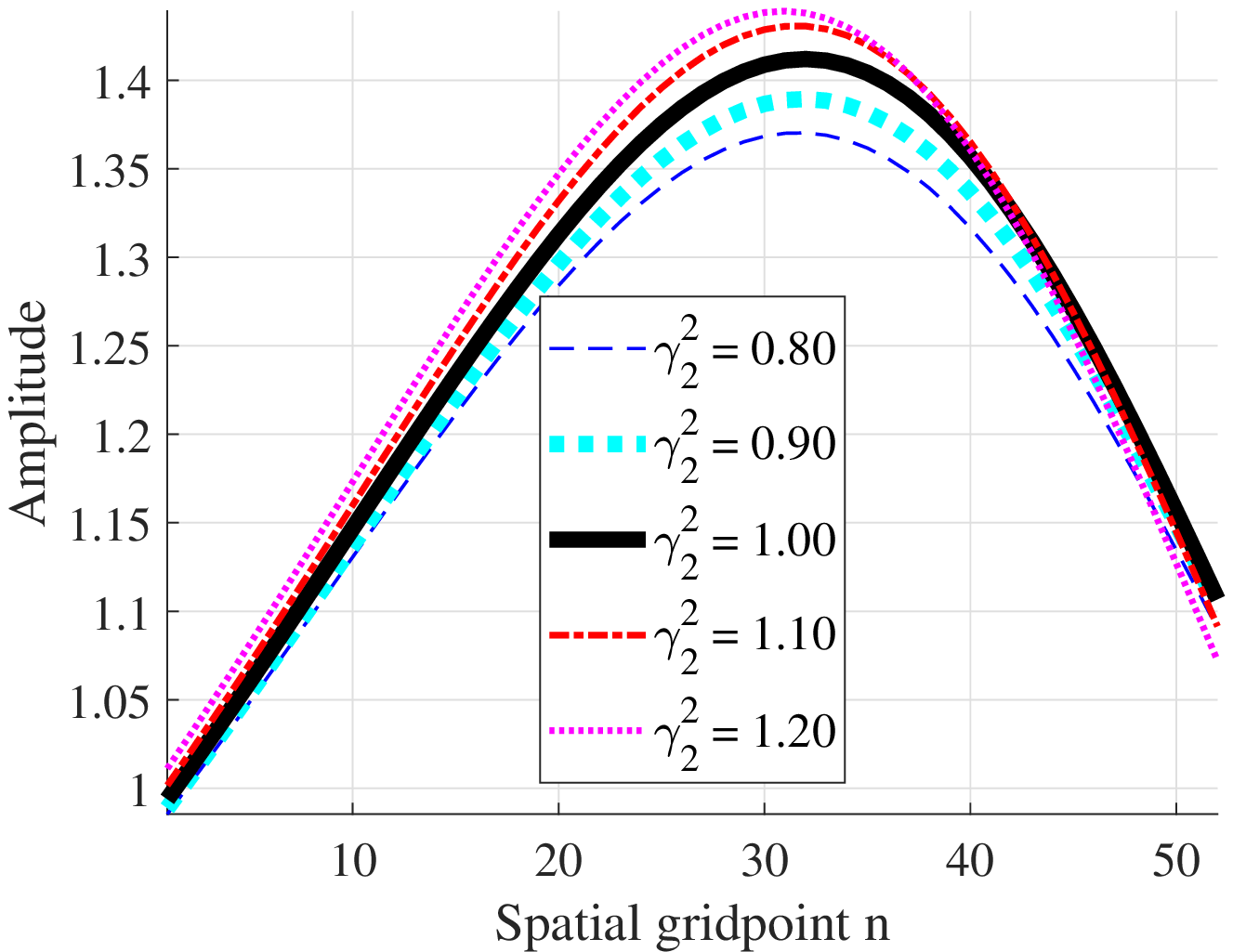}
\includegraphics[width=0.49\textwidth]{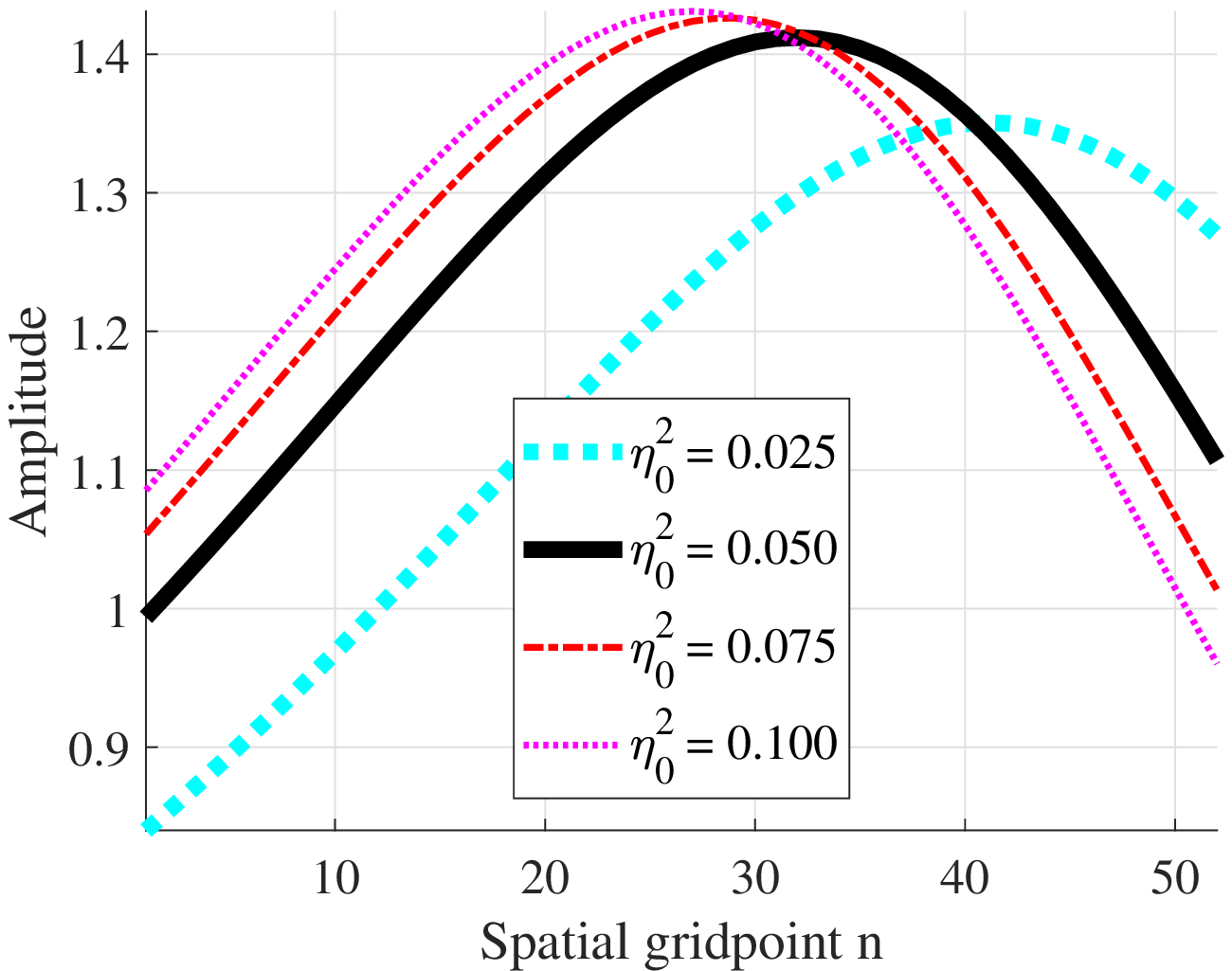}\\
\includegraphics[width=0.49\textwidth]{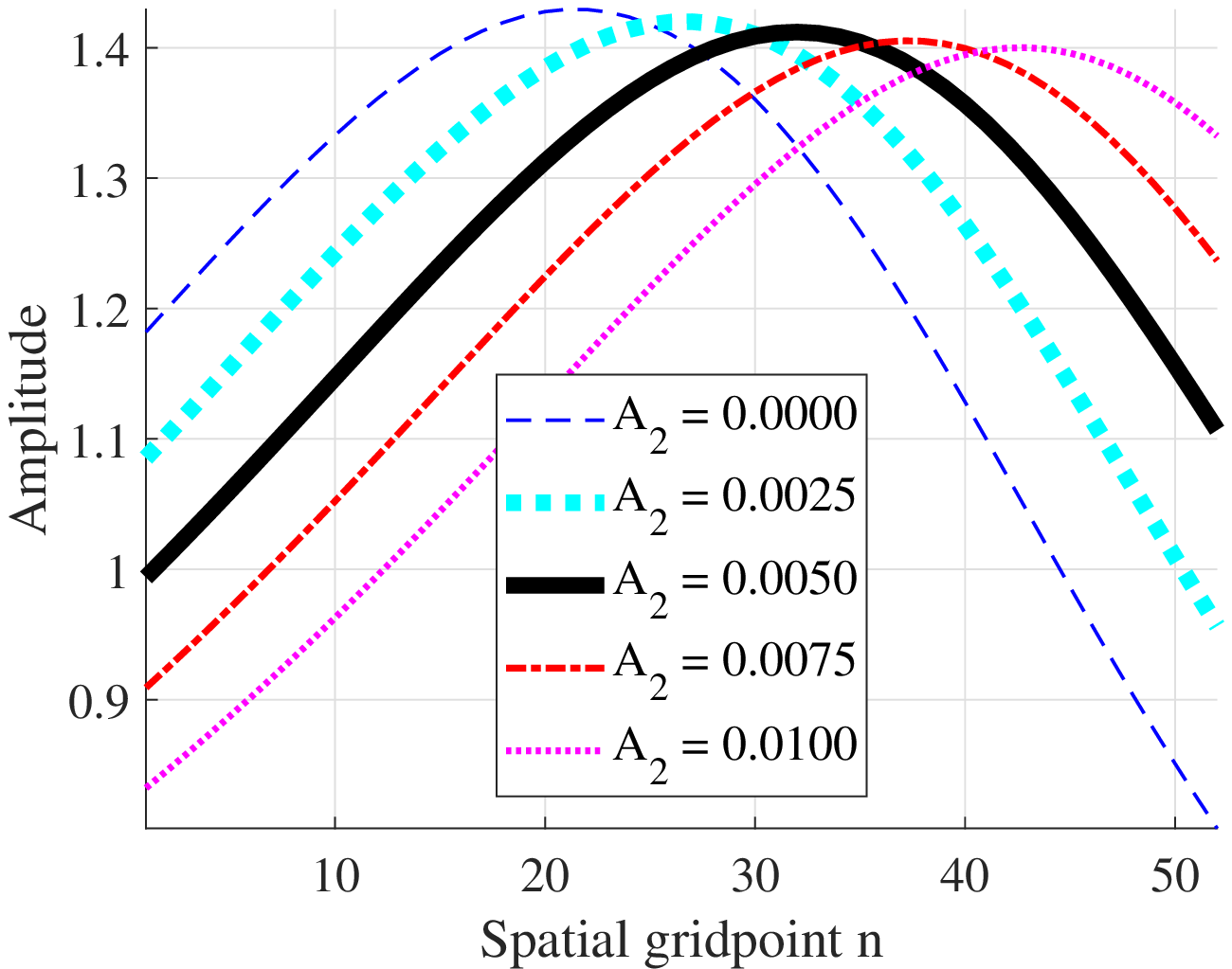}
\includegraphics[width=0.49\textwidth]{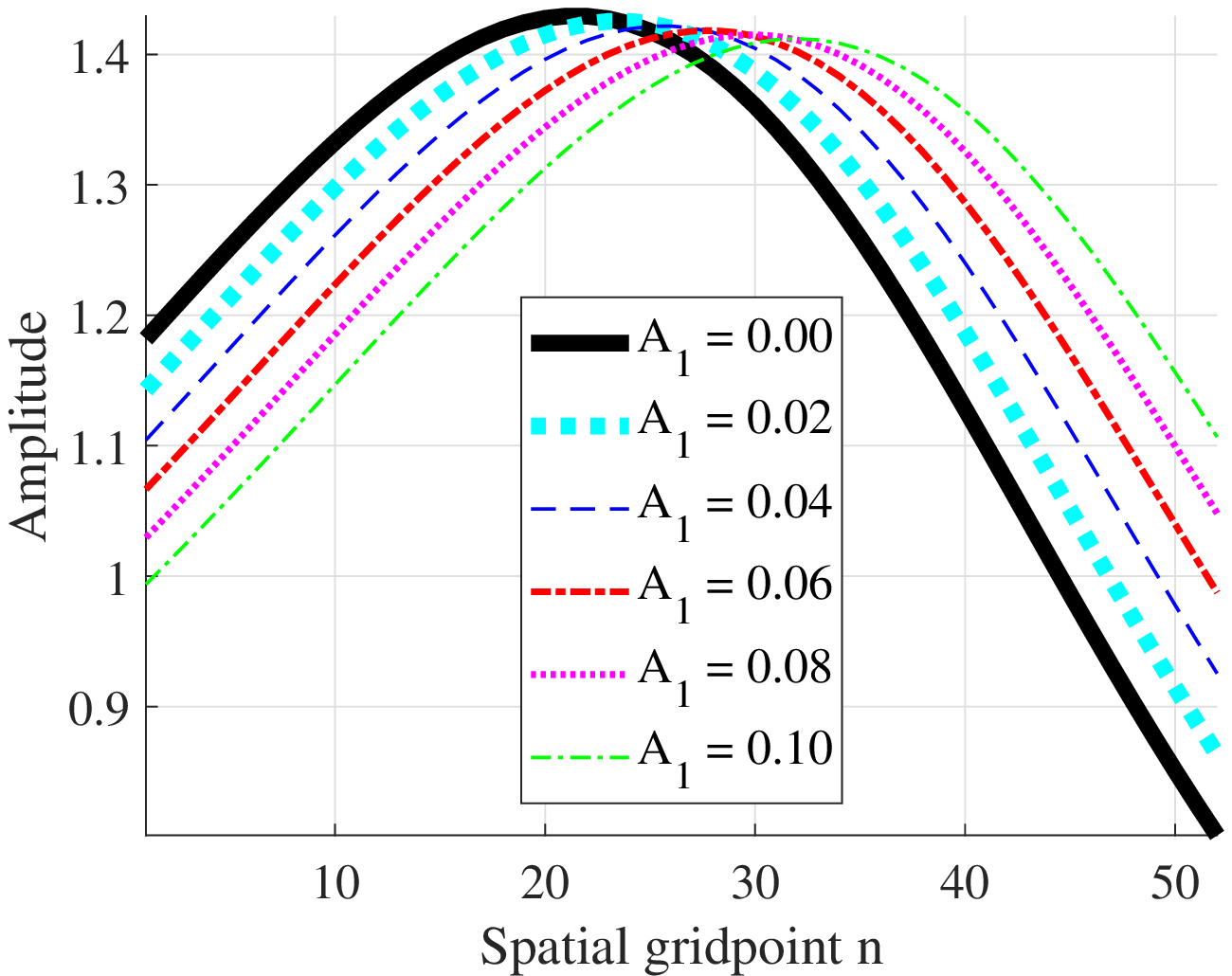}
\caption{A close-up view of the waveprofile peak at $T=250$ from $n=1475$ to $n=1526$ at various parameter values.
Top-left panel -- influence of  $\gamma_2^2$  (at $\gamma_2^2=1$ the sound velocity in myelin is the same as in the regular improved Heimburg-Jackson model); top-right panel -- influence of $\eta_0^2$; bottom-left panel -- influence of $A_2$; bottom-right panel -- influence of $A_1$ (here $A_1=0$ denotes a case where there is no coupling between iHJ and myelin sheath). }
\label{fig6}
\end{figure}

For a closer look we zoom-in to the peak of the left-propagating pulse in Fig.~\ref{fig5} at $T=250$ and try to understand how does changing the parameters $\gamma_{2}^{2},\eta_{0}^{2},A_1$ and $A_2$ alter the evolution of the wave-profile under investigation in Fig.~\ref{fig6}. \\
\textbf{Parameter $\gamma_2^2$} -- this parameter controls, in essence, the influence of the field equation for the myelin sheath on propagating of energy: faster than the lipid bilayer ($\gamma_2^2>\gamma_0^2$) or slower ($\gamma_2^2<\gamma_0^2$). 
Increasing the parameter $\gamma_2^2$ speeds up the wave profile slightly and also leads to a slightly higher amplitude at $T=250$ (see Fig.~\ref{fig6}). \\
\textbf{Parameter $\eta_0^2$} -- increasing this parameter increases the velocity and amplitude of the pulse (Fig.~\ref{fig6} top-right panel).\\
\textbf{Parameter $A_1$} -- increasing this parameter reduces the velocity and amplitude of the pulse (Fig.~\ref{fig6} bottom-left panel).\\
\textbf{Parameter $A_2$} -- this parameter determines the strength of coupling between the added field equation for the myelin influence and the macro-scale wave observed from the iHJ equation. Increasing parameter $A_2$ reduces the velocity and amplitude of the pulse (Fig.~\ref{fig6} bottom-right panel). In other words, including the myelin sheath in our model framework reduces the velocity of the mechanical wave under the used parameters. Although it must be noted this is a result of preliminary numerical analysis and would need experimental verification in future studies. 

%

\section{Conclusions and discussion}

To sum up, we started with a brief overview of the neuron structure and noted the importance of myelin sheath for the correct functioning of some types of neurons. Several experimental observations have shown that the electrical nerve signals are accompanied by mechanical, thermal, and chemical changes and there exist many mathematical models for accounting for the observed effects. We opted to use one of the outlined models (the coupled model) which included the description of the action potential $Z$ (by making use of the FitzHugh-Nagumo model), the longitudinal mechanical wave $U$ in the axon wall (by making use of the improved Heimburg-Jackson model), the pressure wave $P$ in the axoplasm (by making use of a modified wave equation), the temperature changes $\Theta$ (by making use of the modified heat equation) and the endo-- or exothermal effects from abstracted chemical reactions taking into account through internal variables \cite{Raamat2021}. All these building blocks were tied together through coupling forces to form a wave ensemble that represented in this model the whole nerve pulse. However, we left aside most of the included effects in the coupled model and focused on considering only the mechanical wave in the biomembrane in order to modify the relevant governing equation to include the influence of the myelin sheath. For that purpose, an additional wave-type  equation was coupled with the improved Heimburg-Jackson model which represented the influence of the myelin sheath as a field that modulates the observed wave at macro-scale. The resulting governing equations were analyzed by making use of the analytical and numerical methods. Significant attention was directed to the dispersive and dissipative properties of the modified model system followed by a preliminary numerical analysis demonstrating the influence of the added parameters on the evolution of the solutions of the modified system. 

The main conclusion which can be drawn from the results of the numerical example and the observations done during the dispersion analysis is that accounting for the myelin sheath in our model for the mechanical wave on axons reduces the propagation velocity of the long-wavelength (i.e., low frequency) mechanical disturbances along the axon. 

This means that the improved Heimburg-Jackson equation used in the coupled model was successfully expanded to account for the myelin sheath. This is the first step, because as noted earlier, the coupled model accounted for many other effects for the nerve pulse ensemble than only the mechanical wave in the biomembrane, meaning that the next steps would be to consider what does the addition of the myelin sheath mean for the other constituents of the coupled model. 

For example, taking into account the myelin sheath in the coupled model should affect the electrical signal in the nerve pulse ensemble significantly. 
Under the HH paradigm, the mechanism of the myelination increases the propagation velocity of the AP signal in axons and the prevalent explanation of this phenomenon in literature is capacitative in nature. As the capacity of the axon changes significantly between the myelinated and unmyelinated sections, this causes, in a nutshell, the electrical signal to `jump' between the nodes of Ranvier propagating much faster than it would in the case of an unmyelinated axon. It is typically noted that diffusion on its own is not fast enough to explain the much faster propagation of the AP in the myelinated axons and for that reason an alternate mechanism is needed. Interestingly, the propagation velocity of the mechanical wave in a lipid bilayer happens to be in the right velocity range, to significantly speed up the nerve pulse velocity, if it is capable of triggering ion channels in the nodes of Ranvier. It has been estimated that the velocity of the longitudinal density change in the biomembrane could be, very roughly, 175 m/s \cite{Heimburg2005}. This is much faster than the AP in the unmyelinated axon (very roughly 1 m/s) and somewhat faster than the typical AP velocity range in myelinated axons (roughly 10 to 100 m/s). 

Considering the noted characteristics it is possible to outline an alternative path to increased velocity of the AP in the myelinated axons.
The alternative proposal is motivated by the AP generation through flexoelectricity \cite{Mussel2019,Chen2019}. The main point is that the mechanical wave in biomembrane is typically much faster than AP in an unmyelinated axon. The velocity range for an LW in biomembrane is in the right range to be a plausible mechanism for carrying part of the nerve signal between Ranvier nodes much faster than the diffusive processes are capable of. However, such a statement is speculative. The proposed mechanism could be the following:\\
\textbf{(i)} the AP generates the mechanical wave during its formation in the axon hillock, which propagates through the myelinated section(s) of the axon towards the next node of Ranvier;\\
\textbf{(ii)} when the mechanical wave reaches the node of Ranvier the mechanical deformations could be sufficient to exceed the threshold of some ion channels (either voltage-gated, by changing the potential through the flexoelectric effect or mechanosensitive variants) leading to changes of the shape and velocity of the AP. This is the speculative part as although it has been demonstrated in the literature \cite{Mussel2019,Chen2019} that the voltage changes from flexoelectricity could be sufficient, there are some open questions, namely: \textbf{(a)} is there some kind of focusing of mechanical energy into the biomembrane segment in the unmyelinated section, \textbf{(b)} what effects are important and how do they interact with the mechanical wave and the electrical signal (flexoelectricity, ion channels, chemical reactions, pressure, temperature, etc), \textbf{(c)} how significant are the dissipative effects for the mechanical wave;\\
\textbf{(iii)} in the node of Ranvier the renewed AP amplifies the mechanical wave and the process could be repeated `jumping' the AP to the next Ranvier node faster than it should happen otherwise.

It should be emphasized, that the outlined alternative possibility for speeding up the AP in myelinated axons does not need to be exclusive but could co-exist with the capacitative mechanism. Because some ion channels are mechanically sensitive but simultaneously, the changes in the lipid structure curvature could change the membrane surface potential \cite{Mussel2019,Chen2019} which could, in theory, also trigger voltage-gated ion channels if the surface potential changes are sufficient for that. On the other hand, if the electrical signal changes through the capacitative processes it should also change the curvature of the membrane which, in turn, could affect or generate the mechanical wave in the biomembrane. 

Incorporating the noted effects into the coupled model is subject for further studies, as in the present paper the main focus was figuring out how does the inclusion of the myelin sheath for the mechanical wave affect its evolution as it propagates along the axon. Some open questions remain but the first significant step in that direction has been taken. 

\section*{Acknowledgments} This research was supported by the Estonian Research Council (IUT 33-24, PUG 1227). Jüri Engelbrecht acknowledges the support from the Estonian Academy of Sciences.


\end{document}